\documentclass[aip,jcp,reprint,superscriptaddress,noshowkeys]{revtex4-2}
\usepackage{graphicx,dcolumn,bm,microtype,multirow,amscd,amsmath,amssymb,amsfonts,mathtools,physics,wrapfig,siunitx}

\usepackage[version=4]{mhchem}
\usepackage[dvipsnames]{xcolor}
\usepackage[normalem]{ulem}
\usepackage[utf8]{inputenc}

\usepackage{hyperref}
\hypersetup{
    colorlinks,
    linkcolor={red!50!black},
    citecolor={red!70!black},
    urlcolor={red!80!black}
  }

\usepackage[normalem]{ulem}
\newcommand{\red}[1]{\textcolor{black}{#1}}

\newcommand{\LCPQ}{Laboratoire de Chimie et Physique Quantiques (UMR 5626), Universit\'e de Toulouse, CNRS, France}
\begin{document}	

\title{Parquet theory for molecular systems: Formalism and static kernel parquet approximation}

\author{Antoine \surname{Marie}}
%	\email{amarie@irsamc.ups-tlse.fr}
	\affiliation{\LCPQ}
\author{Pierre-Fran\c{c}ois \surname{Loos}}
	\email[Corresponding author: ]{loos@irsamc.ups-tlse.fr}
	\affiliation{\LCPQ}
        
\begin{abstract}
The $GW$ approximation has become a method of choice for predicting quasiparticle properties in solids and large molecular systems, owing to its favorable accuracy-cost balance.
However, its accuracy is the result of a fortuitous cancellation of vertex corrections in the polarizability and self-energy.
Hence, when attempting to go beyond $GW$ through inclusion of vertex corrections, the accuracy can deteriorate if this delicate balance is disrupted.
In this work, we explore an alternative route that theoretically goes beyond $GW$: the parquet formalism.
Unlike approaches that focus on a single correlation channel, such as the electron-hole channel in $GW$ or the particle-particle channel in $T$-matrix theory, parquet theory treats all two-body scattering channels on an equal footing.
We present the formal structure of the parquet equations, which couple the one-body Green's function, the self-energy, and the two-body vertex.
We discuss the approximations necessary to solve this set of equations, the advantages and limitations of this approach, outline its implementation for molecular systems, and assess its accuracy for principal ionization potentials of small molecular systems.
%\bigskip
%\begin{center}
%	\boxed{\includegraphics[width=0.5\linewidth]{TOC}}
%\end{center}
%\bigskip
\end{abstract}

\maketitle

%=================================================================%
\section{Introduction}
\label{sec:intro}
%=================================================================%

The success of the $GW$ approximation~\cite{Hedin_1965} is largely attributed to its favorable balance between computational cost and accuracy.\cite{Aryasetiawan_1998,Reining_2018,Golze_2019,Marie_2024a}
With modern implementations scaling as $\order*{K^3}$ or $\order*{K^4}$ using various algorithmic techniques,\cite{Neuhauser_2014,Govoni_2015,Vlcek_2017,Wilhelm_2018,DelBen_2019,Forster_2020,Kaltak_2020,Forster_2021,Duchemin_2021,Wilhelm_2021,Forster_2022,Yu_2022,Tolle_2024} where $K$ is the size of the one-body basis set, $GW$ has become the method of choice for predicting band structures in solids and ionization potentials (IPs) in large molecular systems.\cite{Strinati_1980,Hybertsen_1985,Hybertsen_1986,Godby_1988,Schone_1998,Onida_1995,Ishii_2001,Ishii_2002,Rohlfing_1995,Grossman_2001,Tiago_2003,Tiago_2006,Tiago_2008,Rostgaard_2010,Blase_2011b,Faber_2011,Ke_2011,Bruneval_2012,Korzdorfer_2012,Umari_2013,Brooks_2020,Graml_2024}
This accuracy is now understood to arise from a fortunate cancellation of errors.\cite{Forster_2024} 

Hedin's equations offer a formal framework to go beyond $GW$ through the systematic inclusion of vertex corrections in the polarizability and the self-energy.\cite{Shirley_1996,DelSol_1994,Schindlmayr_1998,Morris_2007,Shishkin_2007b,Romaniello_2009a,Romaniello_2012,Gruneis_2014,Hung_2017,Maggio_2017b,Wang_2021a,Mejuto-Zaera_2022,Forster_2022a,Weng_2023,Wen_2024,Bruneval_2024,Forster_2024,Forster_2025}
In principle, these vertex corrections should improve the accuracy of quasiparticle properties, particularly in regimes where $GW$ is known to struggle. 
However, in practice, such refinements are often delicate: they may destroy the beneficial error cancellation intrinsic to $GW$, sometimes worsening results rather than improving them.\cite{Lewis_2019,Bruneval_2024,Forster_2024,Forster_2025}
Despite its practical success, $GW$ remains limited in systems with strong correlation, where its assumptions break down.\cite{Verdozzi_1995,DiSabatino_2016,Tomczak_2017,Dvorak_2019a,Dvorak_2019b,DiSabatino_2022,DiSabatino_2023,Orlando_2023b,Ammar_2024}

In light of these limitations, this work explores an alternative strategy: the parquet formalism.\cite{DeDominicis_1964a,DeDominicis_1964b}
Unlike Hedin's equations and the $GW$ approximation, which emphasize the electron-hole (eh) correlation channel, or the particle-particle (pp) $T$-matrix approximation,\cite{Baym_1961,Danielewicz_1984a,Bethe_1957,FetterBook,Marie_2024b} which focuses on the pp correlation channel, the parquet approach seeks to treat all two-body correlation channels on an equal footing.\footnote{Note that in this context particle means either electron or hole.}
Previous works have shown that the relative importance of these channels in finite molecular systems is not obvious, motivating a more balanced treatment.\cite{Zhang_2017,Monino_2023,Marie_2024}
%Unlike standard perturbative approaches, the parquet formalism is well-suited to systems with stronger electronic correlations, where the delicate balance between interaction and kinetic energy makes simple perturbative expansions unreliable.
While several attempts have been made to combine multiple channels in an \textit{ad hoc} manner,\cite{Liebsch_1981,Springer_1998,Zhukov_2005,Romaniello_2012,Nabok_2021} these approaches often lack formal foundations.
By contrast, the parquet formalism provides a rigorous framework for such a treatment by classifying diagrammatic contributions into three correlation channels.

Parquet theory consists of a set of coupled equations that interrelate the one-body Green's function, the self-energy, and the two-body vertex function.
Specifically, the parquet equations combine three core ingredients: the Dyson-Schwinger equation, the Bethe-Salpeter equation in each of the three correlation channels, and the parquet decomposition of the vertex function, which classifies the two-body diagrams into irreducible and three distinct types of reducible diagrams (see below).
This structure enables the self-consistent determination of both one- and two-body quantities if the two-body fully irreducible vertex is known or suitably approximated. 
In practice, the so-called parquet \textit{approximation} is defined by replacing the fully irreducible vertex with the antisymmetrized bare Coulomb interaction.

The parquet formalism was originally introduced in the 1960s by de Dominicis and Martin,\cite{DeDominicis_1964a,DeDominicis_1964b} and later revisited by Bickers and co-workers in the context of the fluctuation-exchange (FLEX) approximation.\cite{Bickers_1989a,Bickers_1989b,Bickers_1991,Serene_1991,Esirgen_1997,Bickers_2004}
This latter combines contributions from every two-body correlation channel (without double counting) to construct the self-energy, and can be viewed as a non-self-consistent variant of the parquet approximation.
More recently, the parquet formalism has attracted renewed interest in the condensed matter community (see also Ref.~\onlinecite{Bergli_2011} for an application in nuclear physics), particularly due to advances in algorithmic developments that have enabled its application to strongly correlated systems such as the Hubbard model at larger interaction strengths.\cite{Yang_2009,Tam_2013,Pudleiner_2019,Li_2019,Krien_2022}
Nonetheless, even with modern computational resources, applying the parquet formalism to larger Hubbard lattices remains challenging due to its significant memory requirements, prompting the development of dedicated numerical schemes.\cite{Li_2016,Eckhardt_2020,Wallerberger_2021,Rohshap_2025}

A central advantage of the parquet approach is that it respects the essential many-body symmetries. 
Crossing symmetry, ensuring that exchange between different correlation channels is properly accounted for, is preserved by construction. 
The parquet scheme also satisfies the Pauli exclusion principle, unlike, for example, the $GW$ approximation, which is known to suffer from self-screening errors.\cite{Nelson_2007,Romaniello_2009a,Aryasetiawan_2012,Wetherell_2018}
By contrast, the $GW$ approximation, and, more generally, Baym-Kadanoff conserving schemes,\cite{Baym_1961,Baym_1962} are designed to enforce conservation laws for one-body quantities such as particle number and energy \red{when they are performed fully self-consistently}.
The parquet formalism sacrifices some conservation properties but gains a systematic, self-consistent treatment of two-particle scattering processes.

In summary, the parquet formalism offers a promising, albeit computationally demanding, route to go beyond $GW$ in a controlled and symmetric manner. 
It is particularly well-suited to cases where different types of correlation are intertwined and cannot be captured by a single correlation channel.
To the best of our knowledge, the parquet formalism has not yet been applied in the context of quantum chemistry.
The goal of this work is to begin exploring this possibility.
The remainder of this manuscript is organized as follows.
Section~\ref{sec:reducibility} introduces the concept of diagrammatic one- and two-body reducibility, which sets the stage for the parquet equations presented in Sec.~\ref{sec:parquet_set}. 
In Sec.~\ref{sec:parquet_expr}, we provide the corresponding expressions in frequency space and a spin-orbital basis, while Sec.~\ref{sec:algorithm} discusses the associated algorithm in detail.
Computational details and numerical results are reported in Secs.~\ref{sec:comp_det} and~\ref{sec:results}, respectively.
Finally, conclusions are drawn in Sec.~\ref{sec:conclusion}.

%=================================================================%
\section{One- and two-body reducibility}
\label{sec:reducibility}
%=================================================================%

%=================================================================%
\subsection{One-body reducibility}
\label{sec:one-reducibility}
%=================================================================%

The time-ordered one-body Green's function, also known as one-body propagator, is defined as
\begin{equation}
  \label{eq:g1}
  G(11') = (-\ii)\mel*{\Psi_0^N}{\Hat{T}\mqty[\hpsi(1) \hpsid(1')]}{\Psi_0^N},
\end{equation}
where $\ket*{\Psi_0^N}$ is the exact $N$-electron ground state wave function.
The time-ordering operator $\hT$ acts as
\begin{equation}
\begin{split}
  \Hat{T}\mqty[\hpsi(1) \hpsid(1')] 
  & = \Theta(t_{1} - t_{1'}) \hpsi(1) \hpsid(1') 
  \\
  & - \Theta(t_{1'} - t_{1}) \hpsid(1') \hpsi(1),
\end{split}
\end{equation}
where $\Theta(t)$ is the Heaviside step function, while $\hpsi(1)$ and $\hpsid(1')$ represent second-quantized annihilation and creation field operators in the Heisenberg picture.\cite{Martin_2016}
Here, $1$ is a space-spin-time composite variable $(1)=(\vb{x}_1,t_1)=(\vb{r}_1,\sigma_1,t_1)$.

\begin{figure}%[t!]
  \includegraphics[width=0.8\linewidth]{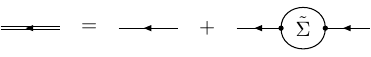}
  \caption{Diagrammatic representation of the full one-body vertex, as defined in Eq.~\eqref{eq:tilde_Sig}.\label{fig:fig1}}
\end{figure}

The one-body propagator can be computed analytically if the Hamiltonian is restricted to its one-body part and is referred to as the independent-particle propagator $G_0$.
Hence, it is natural to compute $G$ perturbatively and to decompose it as
\begin{equation} \label{eq:tilde_Sig}
  G(11') = G_0(11') + G_0(12) \tilde{\Sigma}(22') G_0(2'1'),
\end{equation}
in terms of the full one-body vertex $\tilde{\Sigma}$.
Note that the integration over repeated composite indices is assumed throughout this manuscript.
This equation is represented diagrammatically in Fig.~\ref{fig:fig1}, where the single- and double-line arrows represent $G_0$ and $G$, respectively.\cite{MattuckBook}

%%%  FIG 1  %%% 
%%% %%% %%% %%%

%%%  FIG 2  %%% 
\begin{figure}[b!]
  \includegraphics[width=\linewidth]{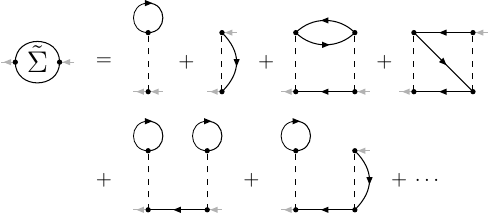}
  \caption{Perturbation expansion of the full one-body vertex $\tilde{\Sigma}$.\label{fig:fig2}}
\end{figure}
%%% %%% %%% %%%

Hence, $G_0$ corresponds to the zeroth-order term of the expansion while $\tilde{\Sigma}$ admits a perturbation expansion in terms of the Coulomb interaction.
The first terms of this expansion are represented diagrammatically in Fig.~\ref{fig:fig2}, where the Coulomb interaction is drawn as a dashed line.\cite{Martin_2016}
The expression corresponding to the first two terms on the right-hand side is
\begin{equation}
  \label{eq:first_order_sigma}
  \Sigma^{(1)} = - \ii v(1 2;1' 2') G(2' 2^+) + \ii v(1 2;2' 1') G(2' 2^+),
\end{equation}
where 
\begin{equation}
  v(1 2; 1' 2') = \delta(11') \frac{\delta(t_1 - t_2)}{\left|\br_1-\br_2\right|} \delta(12),
\end{equation}
and $2^+$ means that a positive infinitesimal shift has been added to the associated time variable.
They are of first order in the Coulomb interaction.
Among the second-order diagrams, the two on the bottom line resemble first-order diagrams glued together with a propagator.
Diagrammatically, they are said to be one-body reducible, \ie, they can be separated into two parts by cutting one propagator line.
On the other hand, the two remaining second-order diagrams do not fulfil this condition and are said to be irreducible.
This classification can be pursued for higher-order diagrams, and the number of irreducible diagrams becomes exponentially smaller than the number of reducible ones with increasing order.

\begin{figure}%[t!]
  \includegraphics[width=0.8\linewidth]{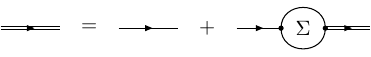}
  \caption{Diagrammatic representation of the Dyson equation, as defined in Eq.~\eqref{eq:dyson_eq}\label{fig:fig3}}
\end{figure}

Hence, it is natural to try to only compute some  selected irreducible diagrams and generate the corresponding reducible diagrams in some systematic way.
The Dyson equation 
\begin{equation}
  \label{eq:dyson_eq}
  G(11') = G_0(11') + G_0(12) \Sigma(22') G(2'1'),
\end{equation}
which expresses $G$ in terms of the \textit{irreducible} one-body vertex $\Sigma$ (also known as the self-energy), which contains every irreducible diagram of $\tilde{\Sigma}$, does exactly this.
It is represented in Fig.~\ref{fig:fig3}.

%%%  FIG 3  %%% 
%%% %%% %%% %%%

The Dyson equation also incorporates every reducible diagram.
This becomes more evident by injecting the expression of $G$ into the right-hand side to produce the following perturbation expansion
\begin{multline}
  \label{eq:dyson_eq_expanded}
  G(11') = G_0(11') + G_0(12) \Sigma(22') G_0(2'1') \\
  + G_0(12) \Sigma(23) G_0(33')  \Sigma(3'2') G_0(2'1') + \cdots.
\end{multline}
Indeed, for a given approximation of $\Sigma$, the Dyson equation generates an infinite number of reducible diagrams during the computation of $G$.
For example, if $G$ is computed through the Dyson equation with $\Sigma = \Sigma^{(1)} + \Sigma^{(2)}$ where 
\begin{equation}
  \label{eq:second_order_self_nrj}
  \begin{split}
    \Sigma^{(2)} &= v(12;3'2') G(3'3) G(4'2) G(2'4) v(34;1'4') 
    \\
                 &- v(12;3'2') G(3'3) G(4'2) G(2'4) v(34;4'1')   
  \end{split}
\end{equation}
represents the two second-order irreducible diagrams of Fig.~\ref{fig:fig2}, then the resulting one-body Green's function is exact up to second order.
This is because the Dyson equation automatically generates the missing reducible second-order contributions, together with an infinite number of reducible terms of arbitrary order.
Physically, the self-energy $\Sigma$ represents all possible interactions experienced by a propagating particle.
This non-local potential encompasses Hartree (H), exchange (x), and correlation (c) effects. 

%=================================================================%
\subsection{Two-body reducibility}
\label{sec:two-reducibility}
%=================================================================%

\begin{figure}%[b!]
  \includegraphics[width=0.8\linewidth]{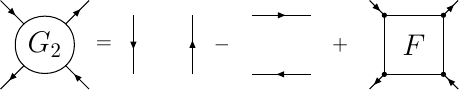}
  \caption{Diagrammatic representation of the full two-body vertex, as defined in Eq.~\eqref{eq:G2_F}.}
  \label{fig:fig4}
\end{figure}

The time-ordered two-body Green's function is defined as
\begin{equation}
  \label{eq:G2}
  G_2(12;1'2') = (-\ii)^2\mel*{\Psi_0^N}{\hT[\hpsi(1)\hpsi(2)\hpsid(2')\hpsid(1')]}{\Psi_0^N}.
\end{equation}
The full two-body vertex $F$ can be defined similarly to the one-body vertex [see Eq.~\eqref{eq:tilde_Sig}] through its relationship with the two-body propagator $G_2$
\begin{multline}
  \label{eq:G2_F}
  G_2(12;34) = G(13) G(24) - G(14) G(23) \\
  - G(11') G(3'3) F(1'2';3'4') G(4'4) G(22').
\end{multline}
This equation can be represented diagrammatically as in Fig.~\ref{fig:fig4}.
Note that in this case, there are two possibilities for the independent-particle propagation.
Using the symmetries of $G_2$, one can show that the full two-body vertex fulfils the following relation\cite{Bickers_2004}
\begin{equation}
  \label{eq:crossing}
  F(12;34) = -F(21;34) = -F(12;43) = F(21;43),
\end{equation}
known as crossing symmetries.

%%%  FIG 4  %%% 
%%% %%% %%% %%%

\begin{figure}[b!]
  \begin{minipage}{0.65\linewidth}
    \includegraphics[width=0.7\linewidth]{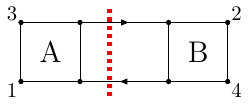}
    \vspace{0.8cm}
    
    \includegraphics[width=0.7\linewidth]{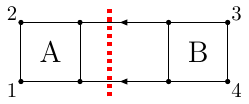}
  \end{minipage}
  \begin{minipage}{0.32\linewidth}
    \includegraphics[width=0.6\linewidth]{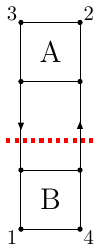} 
  \end{minipage}
  \caption{The three different types of two-particle reducibility lead to three topologically distinct ways of partitioning a diagram into two parts by cutting two propagator lines: eh (top left), $\teh$ (right), and pp (bottom left).}
  \label{fig:fig5}
\end{figure}

The contributions to the full two-body vertex can be classified as either reducible or irreducible diagrams.
In analogy with the one-body case, it is desirable to leverage the power of Dyson equations to generate reducible diagrams starting from irreducible blocks.
However, a key distinction arises compared to the case of $\tilde{\Sigma}$: a two-body diagram can be reducible in multiple ways. 
Figure~\ref{fig:fig5} illustrates the three topologically distinct ways of partitioning a diagram into two parts by cutting two propagator lines, referred to as eh, transversal electron-hole ($\teh$), and pp.

%%%  FIG 5  %%% 
%%% %%% %%% %%%

Using arguments based on the conservation of particle number, it can be shown that a diagram can be reducible only in one of these three channels.
Hence, the full two-body vertex admits the following decomposition\cite{Bickers_2004}
\begin{multline}
  \label{eq:parquet_decomp}
  F(12;34) = \Lambda(12;34) 
  + \Phi^{\eh}(12;34) \\
  + \Phi^{\teh}(12;34) + \Phi^{\pp}(12;34),
\end{multline}
in terms of the fully irreducible two-body vertex $\Lambda$ and the three different reducible two-body vertices $\Phi^{\eh}$, $\Phi^{\teh}$, and $\Phi^{\pp}$.
This decomposition can be written alternatively as
\begin{equation}
  \begin{split}
    \label{eq:}
    F(12;34) &= \Gamma^{\eh}(12;34)  + \Phi^{\eh}(12;34) 
    		\\
             &= \Gamma^{\teh}(12;34) + \Phi^{\teh}(12;34) 
             \\
             &= \Gamma^{\pp}(12;34)  + \Phi^{\pp}(12;34),
  \end{split}
\end{equation}
where the ``irreducible in one channel only'' vertices $\Gamma^{\eh}$, $\Gamma^{\teh}$, and $\Gamma^{\pp}$ have been introduced.
A vertex that is irreducible in one channel is either fully irreducible or reducible in another channel.
Hence, one can deduce the following relationships
\begin{subequations}
  \begin{align}
    \Gamma^{\eh}(12;34)  &= \Lambda(12;34) + \Phi^{\teh}(12;34) + \Phi^{\pp}(12;34), 
    \\
    \Gamma^{\teh}(12;34) &= \Lambda(12;34) + \Phi^{\eh}(12;34)  + \Phi^{\pp}(12;34), 
    \\
    \Gamma^{\pp}(12;34)  &= \Lambda(12;34) + \Phi^{\eh}(12;34)  + \Phi^{\teh}(12;34).
  \end{align}
\end{subequations}

%%%  FIG 6  %%% 
\begin{figure}[b!]
  \vspace{0.5cm}
  \includegraphics[width=\linewidth]{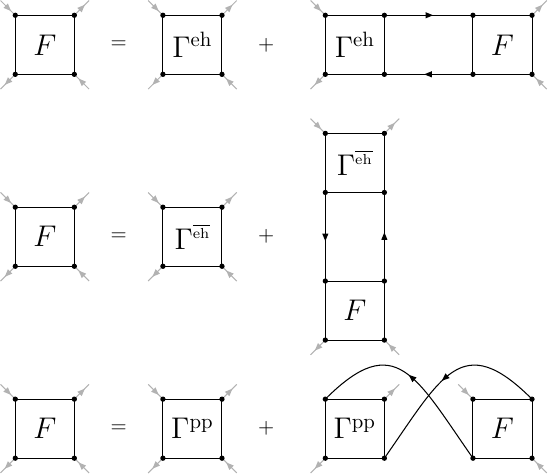}
  \caption{Diagrammatic representation of the three Bethe-Salpeter equations for the full two-body vertex, as defined in Eq.~\eqref{eq:three_f}: eh (top), $\teh$ (center), and pp (bottom).}
  \label{fig:fig6}
\end{figure}
%%% %%% %%% %%%

The next step is to generate reducible diagrams using the irreducible blocks as in the Dyson equation [see Eq.~\eqref{eq:dyson_eq}].
This can be done through three different Bethe-Salpeter equations (BSE) for the full two-body vertex
\begin{subequations}
  \label{eq:three_f}
  \begin{align}
    &\begin{multlined}[0.87\linewidth]
      F(12;34) = \Gamma^{\eh}(12;34) \\
      + \Gamma^{\eh}(13';31') G(1'4')G(2'3') F(4'2;2'4),
    \end{multlined} \\
    &\begin{multlined}[0.87\linewidth]
      F(12;34) = \Gamma^{\teh}(12;34) \\
      - \Gamma^{\teh}(3'2;32') G(1'3')G(2'4') F(14';1'4),
    \end{multlined} \\
    &\begin{multlined}[0.87\linewidth]
      F(12;34) = \Gamma^{\pp}(12;34) \\
      - \frac{1}{2} \Gamma^{\pp}(12;1'2') G(1'3')G(2'4') F(3'4';34).
    \end{multlined}
  \end{align}
\end{subequations}
These three equations are represented diagrammatically in Fig.~\ref{fig:fig6}.
One can immediately see that, by construction, the rightmost terms are reducible in a given channel.

Before ending this section, the crossing symmetries of the irreducible vertices are reported.
The irreducible vertex in the pp channel fulfils the same symmetries as $F$ [see Eq.~\eqref{eq:crossing}].\cite{Bickers_2004}
On the other hand, $\Gamma^{\eh}$ and $\Gamma^{\teh}$ are not crossing symmetric on their own and are related through the following relations\cite{Bickers_2004}
\begin{subequations}
  \begin{align}
  \Gamma^{\eh}(21;34) &= -\Gamma^{\teh}(12;34), \\
  \Gamma^{\eh}(12;43) &= -\Gamma^{\teh}(12;34).
  \end{align}
\end{subequations}

%=================================================================%
\section{Parquet equations}
\label{sec:parquet_set}
%=================================================================%

\begin{figure}[b!]
  \includegraphics[width=\linewidth]{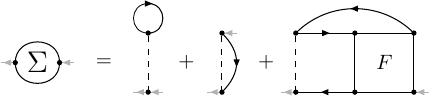}
  \caption{Diagrammatic representation of the self-energy in terms of the full two-body vertex, as defined in Eq.~\eqref{eq:sigma_parquet}.\label{fig:fig7}}
\end{figure}

The one- and two-body vertices can be linked through the equation of motion for $G$, which yields an expression for the self-energy in terms of $G_2$\cite{Martin_2016}
\begin{equation}
  \label{eq:sigma_eom}
  \Sigma(11') = -\ii v(1 2;3' 2') G_2(3' 2';3 2) G^{-1}(31').
\end{equation}
Substituting Eq.~\eqref{eq:G2_F} into this expression leads to
\begin{multline}
  \label{eq:sigma_parquet}
  \Sigma(11') = \Sigma_\text{Hx}(11') \\
  - \ii v(12;3'2') G(3'3) G(4'2) G(2'4) F(34;4'1'),
\end{multline}
where $\Sigma_\text{Hx}$ is the Hartree-Fock (HF) self energy, \ie, the first-order self-energy of Eq.~\eqref{eq:first_order_sigma}.
The remaining term is the desired expression of the correlation self-energy $\Sigma_\text{c}$ in terms of the full two-body vertex.
This decomposition of the self-energy is represented diagrammatically in Fig.~\ref{fig:fig7}.

%%%  FIG 7  %%% 
%%% %%% %%% %%%

The previous discussion laid the groundwork for the self-consistent set of equations known as the parquet equations

\begin{widetext}
  \begin{subequations}
    \label{eq:parquet_eq}
    \begin{align}
      G(11') &= G_0(11') + G_0(12) \Sigma(22') G(2'1'), 
      \\
      \Sigma(11') &= \Sigma_\text{Hx}(11') - \ii v(12;3'2') G(3'3) G(4'2) G(2'4) F(34;4'1'), 
      \\
      F(12;34) &= \Lambda(12;34) + \Phi^{\eh}(12;34) + \Phi^{\teh}(12;34) + \Phi^{\pp}(12;34), 
      \\
      F(12;34) &= \Gamma^{\eh}(12;34)  + \Gamma^{\eh}(13';31') G(1'4')G(2'3') F(4'2;2'4), 
      \\
      F(12;34) &= \Gamma^{\teh}(12;34) - \Gamma^{\teh}(3'2;32') G(1'3')G(2'4') F(14';1'4), 
      \\
      F(12;34) &= \Gamma^{\pp}(12;34)  - \frac{1}{2} \Gamma^{\pp}(12;1'2') G(1'3')G(2'4') F(3'4';34).
    \end{align}
  \end{subequations}
\end{widetext}

Within this set, every quantity can be determined self-consistently except for the fully irreducible vertex $\Lambda$, which must be provided as an input.
The perturbation expansion of $\Lambda$ starts with two first-order terms --- the direct and exchange Coulomb interactions.\cite{Bickers_2004}
A common approximation, known as the parquet \textit{approximation}, assumes
\begin{equation}
  \label{eq:irred_vert_coulomb}
  \Lambda(12;1'2') = -\ii \bar{v}(12;1'2') = -\ii \left[ v(12;1'2') - v(12;2'1') \right].
\end{equation}
This simplification is justified because the next terms in the expansion of $\Lambda$ only appear at fourth order in the Coulomb interaction.\cite{Bickers_2004}
Hence, the corresponding self-energy contributions emerge at fifth order. 
Therefore, solving the parquet equations self-consistently starting from Eq.~\eqref{eq:irred_vert_coulomb} ensures that the self-energy remains exact up to fourth order.

Under the parquet approximation, the self-energy can be written as
\begin{equation}
  \label{eq:sigma_c}
  \Sigma_\text{c}(11') = \Sigma^{(2)}(11') + \Sigma^{\eh}(11') + \Sigma^{\pp}(11'),
\end{equation}
where $\Sigma^{(2)}(11')$ is the part due to the irreducible vertex and is exactly the second-order correlation self-energy
\begin{equation}
  \label{eq:sigma_2}
  \Sigma^{(2)}(11') = - \frac{1}{2} \bar{v}(12;3'2') G(3'3) G(4'2) G(2'4) \bar{v}(34;4'1').
\end{equation}
Using the crossing symmetry, the contribution from $\Phi^{\eh}(12;34)$ and $\Phi^{\teh}(12;34)$ are gathered into 
\begin{equation}
  \label{eq:sigma_eh}
  \Sigma^{\eh}(11') = - \ii \bar{v}(12;3'2') G(3'3) G(4'2) G(2'4) \Phi^{\eh}(34;4'1'),
\end{equation}
while the contribution from the pp reducible kernel is
\begin{equation}
  \label{eq:sigma_pp}
  \Sigma^{\pp}(11') = - \frac{\ii}{2} \bar{v}(12;3'2') G(3'3) G(4'2) G(2'4) \Phi^{\pp}(34;4'1').
\end{equation}

These self-energy expressions highlight that, in the parquet formalism, the resummation in each channel is carried out starting from third order.
This is because the three correlation channels (eh, $\teh$, and pp) share common diagrammatic contributions at second order, and the parquet formalism avoids double counting by construction.
As a result, the expressions for the eh and pp components of the self-energy become more involved than in methods like $GW$ or the $T$-matrix, where the resummation typically starts at second order and is restricted to a single channel.

Note that the $\teh$ reducible vertex does not have to be considered because the self-energy can be expressed only in terms of the eh and pp contributions, as shown above.
The eh and pp reducible vertices can be expressed in closed form by expanding the corresponding BSEs and identifying the reducible contributions, leading to
\begin{subequations}
  \label{eq:reduc_vertex}
  \begin{align}
    \Phi^{\eh}(12;34) &= \Gamma^{\eh}(13';31') L(1'2';3'4') \Gamma^{\eh}(4'2;2'4), 
    \\
    \Phi^{\pp}(12;34) &= - \frac{1}{2} \Gamma^{\pp}(12;1'2') K(1'2';3'4') \Gamma^{\pp}(3'4';34),
  \end{align}
\end{subequations}
where $L$ and $K$ are defined by the eh-BSE and pp-BSE with kernels $\Gamma^{\eh}$ and $\Gamma^{\pp}$, respectively
\begin{subequations}
  \begin{align}
    \label{eq:ehBSE}
    &\begin{multlined}[0.87\linewidth]
      L(12;1'2') = L_0(12;1'2') \\
      + L_0(13';1'3) \Gamma^\eh(34;3'4') L(4'2;42'),
    \end{multlined}
    \\
    \label{eq:ppBSE}
    &\begin{multlined}[0.87\linewidth]
      K(12;1'2') = K_0(12;1'2') \\
      - \frac{1}{2} K(12;44') \Gamma^\pp(44';33') K_0(33';1'2'),
    \end{multlined}
  \end{align}
\end{subequations}
where $L_0(12;1'2') = G(12')G(21')$ and
\begin{equation}
  \label{eq:K0}
  K_0(12;1'2') = \frac{1}{2} \qty[ G(11') G(22') - G(21') G(12') ].
\end{equation}

%=================================================================%
\section{Parquet in practice}
\label{sec:parquet_expr}
%=================================================================%

The solution of the non-linear set of equations \eqref{eq:parquet_eq} under the parquet approximation will now be discussed.
This requires, first, that the time-dependent equations are expressed in frequency space and, then, that the spatial variables are projected in a spin-orbital basis.
\red{Note that, here, the focus is on a solution in terms of spectral representations for each quantity and none of them will be represented on a frequency grid.}
A spin-adapted version of these equations is also provided in the \SupInf.

%%%%%%%%%%%%%%%%%%%%%%%%%%%%%%%%%%%%%%%%%%%%%%%%%%
\subsection{Parquet equations in frequency space}
\label{sec:frequency_space}
%%%%%%%%%%%%%%%%%%%%%%%%%%%%%%%%%%%%%%%%%%%%%%%%%%

The parquet equations involve various 4-point quantities that depend on three time differences because of the time-independent nature of the Hamiltonian.
Hence, the frequency-space equations for the full two-body vertex and the self-energies involve three-frequency quantities.
Note that there are various conventions possible for the three time differences.
The eh-BSE and pp-BSE each have their natural convention.
In the following, each frequency-dependent quantity relies on the eh convention by default, while the P subscript denotes quantities using the pp convention.
These conventions, their link, and the Fourier transform of parquet equations are discussed extensively in the \SupInf.

The three-frequency quantities are particularly challenging to deal with at zero temperature, where the simplifications provided by Matsubara sums are no longer available.
To alleviate this, we first assume that the eh and pp irreducible vertices are static, that is,
\begingroup
\allowdisplaybreaks{}
\begin{subequations}
  \begin{align}
    \Gamma^{\eh}(\bx_{1}\bx_{2};\bx_{3}\bx_{4};\nu,\nu',\omega) &= \Gamma^{\eh}(\bx_{1}\bx_{2};\bx_{3}\bx_{4};\omega=0),
    \\
    \Gamma_{\text{P}}^\pp(\bx_{1}\bx_{2};\bx_{3}\bx_{4};\nu,\nu',\omega) &= \Gamma_{\text{P}}^\pp(\bx_{1}\bx_{2};\bx_{3}\bx_{4};\omega=0),
  \end{align}
\end{subequations}%
\endgroup
or, alternatively, one can say that the eh and pp irreducible vertices are assumed to be instantaneous effective interactions.

Hence, the reducible vertices [see Eqs.~\eqref{eq:reduc_vertex}] depend only on their respective natural bosonic frequencies
% \begin{align}
%   \Phi^{\eh}(\bx_{1}\bx_{2};\bx_{3}\bx_{4};\nu,\nu',\omega) &= \Phi^{\eh}(\bx_{1}\bx_{2};\bx_{3}\bx_{4};\omega),
%   \\
%   \Phi_{\text{P}}^\pp(\bx_{1}\bx_{2};\bx_{3}\bx_{4};\nu,\nu',\omega) &= \Phi_{\text{P}}^\pp(\bx_{1}\bx_{2};\bx_{3}\bx_{4};\omega),
% \end{align}
and their expression are given by
\begin{multline}
  \label{eq:eh_reduc_vertex}
  \Phi^{\eh}(\bx_{1}\bx_{2};\bx_{3}\bx_{4};\omega) = \Gamma^{\eh}(\bx_{1}\bx_{3'};\bx_{3}\bx_{1'}) \\
  \times L(\bx_{1'}\bx_{2'};\bx_{3'}\bx_{4'};\omega) \Gamma^{\eh}(\bx_{4'}\bx_{2};\bx_{2'}\bx_{4}),    
\end{multline}
and
\begin{multline}
  \label{eq:pp_reduc_vertex}
  \Phi_{\text{P}}^\pp(\bx_{1}\bx_{2};\bx_{3}\bx_{4};\omega) = - \frac{1}{2} \Gamma^{\pp}_{\text{P}}(\bx_{1}\bx_{2};\bx_{1'}\bx_{2'}) \\
  \times K_{\text{P}}(\bx_{1'}\bx_{2'};\bx_{3'}\bx_{4'};\omega) \Gamma^{\pp}_{\text{P}}(\bx_{3'}\bx_{4'};\bx_{3}\bx_{4}),
\end{multline}
where $\omega=0$ has been removed from the static kernels for the sake of conciseness.
This \textit{static kernel approximation} leads to the following expressions for the eh and pp components of the self-energy
\begin{widetext}
  \begin{subequations}
    \begin{align}
      &\begin{multlined}[0.925\textwidth] \Sigma^{\eh}(\bx_1\bx_{1'};\omega) = -\frac{\ii}{(2\pi)^2} \bar{v}(\bx_1\bx_2;\bx_{3'}\bx_{2'}) \\
        \times \int\dd{(\omega_{1}\omega_{2})} G(\bx_{3'}\bx_3;\omega_1) G(\bx_{4'}\bx_2;\omega_2) G(\bx_{2'}\bx_4;\omega - \omega_1 + \omega_2) \Phi^\eh(\bx_3\bx_4;\bx_{4'}\bx_{1'};\omega_{2} - \omega_1),
      \end{multlined} \\
      &\begin{multlined}[0.925\textwidth] \Sigma^{\pp}(\bx_1\bx_{1'};\omega) = -\frac{\ii}{2(2\pi)^2} \bar{v}(\bx_1\bx_2;\bx_{3'}\bx_{2'}) \\
        \times \int\dd{(\omega_{1}\omega_{2})} G(\bx_{3'}\bx_3;\omega_1) G(\bx_{4'}\bx_2;\omega_2) G(\bx_{2'}\bx_4;\omega - \omega_1 + \omega_2) \Phi^\pp_{\text{P}}(\bx_3\bx_4;\bx_{4'}\bx_{1'};\omega_{2} + \omega).
      \end{multlined}
    \end{align}
  \end{subequations}
\end{widetext}
While this approximation greatly simplifies the equations, it is also quite severe and must therefore be carefully evaluated in practice.\cite{Karrasch_2008,Husemann_2012,Vilardi_2017}
Finally, although this may not yet be apparent, the next section will make it clear that this approximation is fully equivalent to the standard static kernel approximation used in the eh-BSE and pp-BSE formalisms.\cite{Strinati_1988,Marie_2025a}

%%%%%%%%%%%%%%%%%%%%%%%%%%%%%%%%%%%%%%%%%%%%%%%%%%
\subsection{Projection in a spin-orbital basis}
\label{sec:parquet_basis}
%%%%%%%%%%%%%%%%%%%%%%%%%%%%%%%%%%%%%%%%%%%%%%%%%%

The aim of self-consistently solving the parquet equations is to determine the one-body propagator.
Assuming a quasiparticle representation, it has the following form in a spin-orbital basis
\begin{equation}
  G_{pq}(\omega) = \sum_i \frac{\delta_{pi}\delta_{qi}}{\omega - \epsilon_i - \ii\eta} + \sum_a \frac{\delta_{pa}\delta_{qa}}{\omega - \epsilon_a + \ii\eta},
\end{equation}
where $\epsilon_p$ are the quasiparticle energies and can be directly linked to the ionization potentials and electron affinities of the system, and $\eta$ is a positive infinitesimal ensuring the correct causal structure.

In this work, the indices $p,q,r,s, \dots$ are used for \red{the $K$} arbitrary orbitals, $i,j,k,l$ label the $O$ occupied orbitals, and $a,b,c,d$ denote the $V$ virtual orbitals.
The indices $n$ and $m$ are occupied-virtual and occupied-occupied/virtual-virtual composite indices, respectively.

The quasiparticle energies can be obtained by solving the Dyson equation [see Eq.~\eqref{eq:dyson_eq}], which in a basis set, takes the form of a frequency-dependent, non-Hermitian mean-field problem, known as the quasiparticle equation
\begin{equation}
  \qty[\bF + \bSig_{\text{c}}(\omega=\epsilon_p) ] \SO{p}(\bx) = \epsilon_p \SO{p}(\bx),
\end{equation}
Here, $\bF$ is the Fock matrix in the spin-orbital basis that includes Hartree-exchange contributions, $\bSig_{\text{c}}(\omega)$ is the frequency-dependent correlation self-energy matrix, and $\SO{p}(\bx)$ are the so-called Dyson orbitals.
Hence, to determine the quasiparticle energies, one must first compute $\bSig_{\text{c}}(\omega)$.
In the parquet framework, the self-energy is given by Eq.~\eqref{eq:sigma_c}, which requires prior computation of the eh and pp reducible vertices.
Sections \ref{sec:eh-red-vertex} and \ref{sec:pp-red-vertex} detail the evaluation of $\Phi^\eh$ and $\Phi^\pp$ in a spin-orbital basis, followed by the final expression for the self-energy in Sec.~\ref{sec:self-energy}.

%///////////////////////////%
\subsubsection{Electron-hole reducible vertex}
\label{sec:eh-red-vertex}
%///////////////////////////%

To compute the eh reducible vertex [see Eq.~\eqref{eq:eh_reduc_vertex}], one must first obtain the corresponding eh propagator $L$.
As shown in the \SupInf, solving the eh-BSE is equivalent to diagonalizing an effective Hamiltonian matrix defined as
\begin{multline} \label{eq:cHeh}
  \mqty(\bA^{\eh} & \bB^{\eh} \\ - (\bB^{\eh})^\dagger & - (\bA^{\eh})^\dagger) \mqty(\bX^{\eh} & (\bY^{\eh})^\dagger \\ \bY^{\eh} & (\bX^{\eh})^\dagger) = \\
  \mqty(\bX^{\eh} & (\bY^{\eh})^\dagger \\ \bY^{\eh} & (\bX^{\eh})^\dagger) \mqty(\boldsymbol{\Omega}^\eh & \boldsymbol{0} \\ \boldsymbol{0} & -\boldsymbol{\Omega}^\eh),
\end{multline}
where the elements of the various blocks are
\begin{subequations} \label{eq:ABeh}
  \begin{align}
    A^{\eh}_{ia,jb} &= (\epsilon_a - \epsilon_i) \delta_{ab}\delta_{ij} + \ii \Gamma^\eh_{ajib}, 
    \\
    B^{\eh}_{ia,bj} &= \ii \Gamma^\eh_{abij}.
  \end{align}
\end{subequations}
Here, $\Gamma^{\eh}_{pqrs}$ is a short notation for $\Gamma^{\eh}_{pqrs}(\omega=0)$.
The eigenvalues correspond to neutral excitation (and deexcitation) energies.
This defines a propagator $L$ from which the eh reducible vertex can be built using Eq.~\eqref{eq:eh_reduc_vertex}, as follows
\begin{equation}
  \Phi^{\eh}_{pqrs}(\omega) = - \sum_{tuvw} (\ii \Gamma^{\eh}_{pvrt}) L_{tuvw}(\omega) (\ii \Gamma^{\eh}_{wqus}),
\end{equation}
which, once the spectral function of $L$ is substituted, reads
\begin{equation}
  \label{eq:eh_effective_interaction}
  \ii \Phi^{\eh}_{pqrs}(\omega) = \sum_{n} \left[ \frac{M^{\eh}_{pr,n} M^{\eh,*}_{sq,n}}{\omega - (\Omega^\eh_n - \ii \eta)} - \frac{M^{\eh,*}_{rp,n} M^{\eh}_{qs,n}}{\omega - (- \Omega^\eh_n + \ii \eta)} \right],
\end{equation}
where the eh generalized effective integrals are
\begin{equation}
  \label{eq:parquet_eh_screened_integrals}
    M^{\eh}_{pq,n} = \sum_{ia} ( \ii \Gamma^{\eh}_{paqi} ) X^{\eh}_{ia,n} + \sum_{ai} ( \ii \Gamma^{\eh}_{piqa} ) Y^{\eh}_{ai,n}.
\end{equation}
Note that we use the same notation as in the $GW$ case, as these can be viewed as generalized $GW$-type effective integrals.
The $GW$ effective integrals are recovered in the limit $\Gamma^{\eh}_{pqrs} = -\ii \braket*{pq}{rs}$.
Finally, the static limit of the reducible kernel is
\begin{equation}
  \label{eq:static_eh_reduc_kernel}
  \ii \Phi^{\eh}_{pqrs} = - \sum_{n} \qty[ \frac{M^{\eh}_{pr,n} M^{\eh,*}_{sq,n}}{\Omega^\eh_n - \ii\eta} + \frac{M^{\eh,*}_{rp,n} M^{\eh}_{qs,n}}{\Omega^\eh_n - \ii\eta} ],
\end{equation}
where $\Phi^{\eh}_{pqrs}$ is a short notation for $\Phi^{\eh}_{pqrs}(\omega=0)$.
\red{Note that, for real orbitals, and if $\eta$ is set to zero, the amplitudes $M^{\eh}_{pq,n}$ and the reducible kernel elements $\ii \Phi^{\eh}_{pqrs}$ are real-valued.}

%///////////////////////////%
\subsubsection{Particle-particle reducible vertex}
\label{sec:pp-red-vertex}
%///////////////////////////%

Likewise, the evaluation of the pp reducible vertex begins with the computation of the corresponding pp propagator [see Eq.~\eqref{eq:pp_reduc_vertex}].
As demonstrated in the \SupInf, solving the pp-BSE is equivalent to diagonalizing an effective Hamiltonian matrix defined as
\begin{multline} \label{eq:cHpp}
  \mqty(\bC^{\pp} & \bB^{\pp} \\ - (\bB^{\pp})^\dagger & -\bD^{\pp}) \mqty(\bX^{\ee} & \bY^{\hh} \\ \bY^{\ee} & \bX^{\hh}) = \\
  \mqty(\bX^{\ee} & \bY^{\hh} \\ \bY^{\ee} & \bX^{\hh}) \mqty(\boldsymbol{\Omega}^{\ee} & \boldsymbol{0} \\ \boldsymbol{0} & -\boldsymbol{\Omega}^{\hh}),
\end{multline}
where the elements of the various blocks are
\begin{subequations} \label{eq:BCDpp}
  \begin{align}
    C^{\pp}_{ab,cd} &= (\epsilon_a + \epsilon_b) \delta_{ac}\delta_{bd} + \ii \Gamma^\pp_{abcd}, 
    \\
    B^{\pp}_{ab,ij} &= + \ii \Gamma^\pp_{abij},
    \\
    D^{\pp}_{ij,kl} &= - (\epsilon_i + \epsilon_j) \delta_{ik}\delta_{jl} + \ii \Gamma^\pp_{ijkl}.
  \end{align}
\end{subequations}
Here, $\Gamma^{\pp}_{pqrs}$ is a short notation for $(\Gamma_{\text{P}}^\pp)_{pqrs}(\omega=0)$.
The eigenvalue matrices $\boldsymbol{\Omega}^{\ee}$ and $\boldsymbol{\Omega}^{\hh}$ correspond to double electron affinities and double ionization potentials, respectively.
This defines a propagator $K$, with elements $K_{\overline{pqrs}}$ in the antisymmetrized \red{two-electron spin-orbital basis set $[\phi_p(\bx_1)\phi_q(\bx_2) - \phi_q(\bx_1)\phi_p(\bx_2)]/\sqrt{2}$,} from which the pp reducible vertex can be built using Eq.~\eqref{eq:pp_reduc_vertex}
\begin{equation} 
  \begin{split}
    (\Phi_{\text{P}}^\pp)_{pqrs}(\omega) &= - \frac{1}{2} \sum_{\substack{t<u \\ v<w}} (\Gamma^{\pp}_{pq\overline{tu}}) K_{\overline{tuvw}}(\omega) (\Gamma^{\pp}_{\overline{vw}rs}) \\
                                         &= - \sum_{\substack{t<u \\ v<w}} (\Gamma^{\pp}_{pqtu}) K_{\overline{tuvw}}(\omega) (\Gamma^{\pp}_{vwrs}) \\
                                         &= + \sum_{\substack{t<u \\ v<w}} (\ii \Gamma^{\pp}_{pqtu}) K_{\overline{tuvw}}(\omega) (\ii \Gamma^{\pp}_{vwrs}),
  \end{split}
\end{equation}
which, once the spectral function of $K$ is substituted, reads
\begin{equation}
  (\ii \Phi^{\pp}_{\text{P}})_{pqrs}(\omega) = \sum_{m} \left[ \frac{M^{\ee}_{pq,m} M^{\ee,*}_{rs,m}}{\omega - (\Omega_m^{\ee}-\ii\eta)} - \frac{M^{\hh,*}_{pq,m} M^{\hh}_{rs,m}}{\omega - (\Omega_m^{\hh}+\ii\eta)} \right],
\end{equation}
where the pp generalized effective integrals are
\begin{subequations}
  \label{eq:parquet_pp_screened_integrals}
  \begin{align}
      M^{\ee}_{pq,m} &= \sum_{a<b} ( \ii\Gamma^{\pp}_{pqab} ) X^{\ee}_{ab,m} + \sum_{i<j} ( \ii\Gamma^{\pp}_{pqij} ) Y^{\ee}_{ij,m}, \\
      M^{\hh}_{pq,m} &= \sum_{i<j} ( \ii\Gamma^{\pp}_{pqij} ) X^{\hh}_{ij,m} + \sum_{a<b} ( \ii\Gamma^{\pp}_{pqab} ) Y^{\hh}_{ab,m}.
  \end{align}
\end{subequations}
As before, we have chosen to use the same notation as in the pp $T$-matrix case, since the pp $T$-matrix effective integrals are recovered in the limit $\Gamma^{\pp}_{pqrs} = -\ii \mel*{pq}{}{rs}$.
Finally, the static limit of the reducible pp kernel is
\begin{equation}
  \label{eq:static_pp_reduc_kernel}
  \ii \Phi^{\pp}_{pqrs} = \sum_{m} \left[ - \frac{M^{\ee}_{pq,m} M^{\ee,*}_{rs,m}}{ \Omega_m^{\ee} - \ii\eta } + \frac{M^{\hh,*}_{pq,m} M^{\hh}_{rs,m}}{ \Omega_m^{\hh} + \ii\eta} \right],
\end{equation}
where $\Phi^{\pp}_{pqrs}$ is a short notation for $(\Phi_{\text{P}}^\pp)_{pqrs}(\omega=0)$.

%///////////////////////////%
\subsubsection{Irreducible vertices}
%///////////////////////////%

Sections \ref{sec:eh-red-vertex} and \ref{sec:pp-red-vertex} have shown that computing the reducible kernels requires solving two BSEs.
These two eigenvalue problems depend on the eh and pp irreducible kernels, which can be expressed as
\begin{subequations}
  \begin{align}
    \Gamma^{\eh}(12;34)  &= -\ii \bar{v}(12;34) - \Phi^{\eh}(12;43) + \Phi^{\pp}(12;34), 
    \\
    \Gamma^{\pp}(12;34)  &= -\ii \bar{v}(12;34) + \Phi^{\eh}(12;34) - \Phi^{\eh}(12;43),
  \end{align}
\end{subequations}
using the crossing symmetry of $\Gamma^{\teh}$ and the parquet approximation.

Once projected in a basis, the eh and pp irreducible kernels read
\begin{subequations} 
\begin{align} 
  \label{eq:update_Gamma_eh}
  \ii \Gamma^{\eh}_{pqrs} & = \mel*{pq}{}{rs} - \ii \Phi^{\eh}_{pqsr} + \ii \Phi^{\pp}_{pqrs},
  \\
  \label{eq:update_Gamma_pp}
  \ii \Gamma^{\pp}_{pqrs} & = \mel*{pq}{}{rs} + \ii \Phi^{\eh}_{pqrs} - \ii \Phi^{\eh}_{pqsr}.
\end{align}
\end{subequations}

%///////////////////////////%
\subsubsection{Self-energy}
\label{sec:self-energy}
%///////////////////////////%

Now that the expressions of the (ir)reducible vertices have been derived and discussed, the last ingredients that must be computed are the three components of the self-energy defined in Eqs.~\eqref{eq:sigma_2},~\eqref{eq:sigma_eh}, and~\eqref{eq:sigma_pp}.
The expression of the second-order self-energy is well known\cite{Suhai_1983,Holleboom_1990,Casida_1989,Casida_1991,SzaboBook,Ortiz_2013,Phillips_2014,Rusakov_2016,Hirata_2015,Backhouse_2021,Backhouse_2020b,Backhouse_2020a} and reads
\begin{equation} \label{eq:Sig_2_pq}
  \begin{split}
    \Sigma_{pq}^{(2)}(\omega) &= \frac{1}{2} \sum_{ija} \frac{\mel*{pa}{}{ji} \mel*{ji}{}{qa}}{\omega - (\epsilon_j + \epsilon_i - \epsilon_a + 3\ii\eta)} \\
                              &+ \frac{1}{2} \sum_{iab} \frac{\mel*{pi}{}{ba} \mel*{ba}{}{qi}}{\omega - (\epsilon_a + \epsilon_b - \epsilon_i - 3\ii\eta)}.
  \end{split}
\end{equation}
The projection in a basis of the eh and pp components of the self-energy yields
\begin{subequations}
  \begin{align}
    &\begin{multlined}[0.85\linewidth] \Sigma_{pq}^{\eh}(\omega) = -\frac{\ii}{(2\pi)^2} \sum_{rst} \mel{pr}{}{st} \int\dd{(\omega_{1}\omega_{2})} G_{ss}(\omega_1) 
    \\
    \times G_{rr}(\omega_2) G_{tt}(\omega - \omega_1 + \omega_2) \Phi^\eh_{strq}(\omega_{2} - \omega_1), \end{multlined}
    \\
    &\begin{multlined}[0.85\linewidth] \Sigma_{pq}^{\pp}(\omega) = -\frac{\ii}{2(2\pi)^2} \sum_{rst} \mel{pr}{}{st} \int\dd{(\omega_{1}\omega_{2})} G_{ss}(\omega_1) 
    \\
    \times G_{rr}(\omega_2) G_{tt}(\omega - \omega_1 + \omega_2) (\Phi^\pp_{\text{P}})_{strq}(\omega_{2} + \omega). \end{multlined}
  \end{align}
\end{subequations}
The closed-form evaluation of these integrals is performed in the \SupInf, and yields
\begingroup
\allowdisplaybreaks{}
\begin{equation} \label{eq:Sig_eh_pq}
  \begin{split}
    &\Sigma_{pq}^{\eh}(\omega) = \\
    &+\sum_{ijan} \frac{\mel{pa}{}{ij}}{\epsilon_a - \epsilon_i - \Omega^\eh_{n} - \ii\eta} \frac{M^{\eh}_{ia,n} M^{\eh,*}_{qj,n}}{\omega - (\epsilon_j - \Omega^\eh_{n} + 2\ii\eta)} 
    \\
    &-\sum_{ijan} \frac{\mel{pa}{}{ij}}{\epsilon_a - \epsilon_i - \Omega^\eh_{n} - \ii\eta} \frac{M^{\eh}_{ia,n} M^{\eh,*}_{qj,n}}{\omega - (\epsilon_i + \epsilon_j - \epsilon_a + 3\ii\eta)}
    \\
    &+ \sum_{ijan} \frac{\mel{pa}{}{ij}}{\epsilon_a - \epsilon_i + \Omega^\eh_{n} - 3\ii \eta} \frac{M^{\eh,*}_{ai,n} M^{\eh}_{jq,n}}{\omega - (\epsilon_i + \epsilon_j - \epsilon_a + 3\ii\eta)} 
    \\
    & + \sum_{ijan} \frac{\mel{pi}{}{aj}}{\epsilon_a - \epsilon_i + \Omega^\eh_{n} - 3\ii\eta} \frac{M^{\eh}_{ai,n} M^{\eh,*}_{qj,n}}{\omega - (\epsilon_j - \Omega^\eh_{n} + 2\ii\eta)} 
    \\
    &+ \sum_{iabn} \frac{\mel{pi}{}{ab}}{\epsilon_a - \epsilon_i - \Omega^\eh_{n} - \ii\eta} \frac{M^{\eh,*}_{ia,n} M^{\eh}_{bq,n}}{\omega - (\epsilon_b + \Omega^\eh_{n} - 2\ii\eta)} 
    \\
    &- \sum_{iabn} \frac{\mel{pi}{}{ab}}{\epsilon_a - \epsilon_i - \Omega^\eh_{n} - \ii\eta} \frac{M^{\eh,*}_{ia,n} M^{\eh}_{bq,n}}{\omega - (\epsilon_a + \epsilon_b - \epsilon_i - 3\ii\eta)} 
    \\
    &+ \sum_{iabn} \frac{\mel{pi}{}{ab}}{\epsilon_a - \epsilon_i + \Omega^\eh_{n} - 3\ii\eta} \frac{M^{\eh}_{ai,n} M^{\eh,*}_{qb,n}}{\omega - (\epsilon_a + \epsilon_b - \epsilon_i - 3\ii\eta)} 
    \\
    & + \sum_{iabn} \frac{\mel{pa}{}{ib}}{\epsilon_a - \epsilon_i + \Omega^\eh_{n} - 3\ii\eta} \frac{M^{\eh,*}_{ai,n} M^{\eh}_{bq,n}}{\omega - (\epsilon_b + \Omega^\eh_{n} - 2\ii \eta)},
  \end{split}
\end{equation}%
\endgroup
for the eh part and
\begingroup
\allowdisplaybreaks{}
\begin{equation} \label{eq:Sig_pp_pq}
  \begin{split}
    &\Sigma_{pq}^{\pp}(\omega) = \\
    &+ \frac{1}{2} \sum_{ijam} \frac{\mel{pa}{}{ij}}{\Omega^\hh_m - \epsilon_i - \epsilon_j - \ii\eta} \frac{M_{ij,m}^{\hh,*} M_{aq,m}^{\hh}}{\omega - (\Omega^\hh_m - \epsilon_a + 2\ii \eta)} 
    \\
    &- \frac{1}{2} \sum_{ijam} \frac{\mel{pa}{}{ij}}{\Omega^\hh_m - \epsilon_i - \epsilon_j - \ii\eta} \frac{M_{ij,m}^{\hh,*} M_{aq,m}^{\hh}}{\omega - (\epsilon_i + \epsilon_j - \epsilon_a + 3\ii\eta)} 
    \\
    &+ \frac{1}{2} \sum_{ijam} \frac{\mel{pa}{}{ij}}{\Omega^\ee_m - \epsilon_i - \epsilon_j - 3\ii\eta} \frac{M_{ij,m}^\ee M_{aq,m}^{\ee,*}}{\omega - (\epsilon_i + \epsilon_j - \epsilon_a + 3\ii\eta)} 
    \\
    &+ \frac{1}{2} \sum_{abcm} \frac{\mel{pa}{}{bc}}{\epsilon_b + \epsilon_c - \Omega^\hh_m - 3\ii \eta} \frac{M_{bc,m}^{\hh,*} M_{aq,m}^{\hh}}{\omega - (\Omega^\hh_m - \epsilon_a + 2\ii \eta)} 
    \\
  	&+ \frac{1}{2} \sum_{iabm} \frac{\mel{pi}{}{ab}}{\epsilon_a + \epsilon_b - \Omega^\ee_m - \ii\eta} \frac{M_{ab,m}^\ee M_{iq,m}^{\ee,*}}{\omega - (\Omega^\ee_m - \epsilon_i - 2\ii \eta)} 
	\\
    &- \frac{1}{2} \sum_{iabm} \frac{\mel{pi}{}{ab}}{\epsilon_a + \epsilon_b - \Omega^\ee_m - \ii\eta} \frac{M_{ab,m}^\ee M_{iq,m}^{\ee,*}}{\omega  - (\epsilon_a + \epsilon_b - \epsilon_i - 3\ii\eta)} 
	\\
    &+ \frac{1}{2} \sum_{iabm} \frac{\mel{pi}{}{ab}}{\epsilon_a + \epsilon_b - \Omega^\hh_m - 3\ii \eta} \frac{M_{ab,m}^{\hh,*} M_{iq,m}^{\hh}}{\omega - (\epsilon_a + \epsilon_b - \epsilon_i - 3\ii\eta)} 
    \\
    &+ \frac{1}{2} \sum_{ijkm} \frac{\mel{pi}{}{jk}}{\Omega^\ee_m - \epsilon_j - \epsilon_k - 3\ii \eta} \frac{M_{jk,m}^\ee M_{iq,m}^{\ee,*}}{\omega - (\Omega^\ee_m - \epsilon_i - 2\ii \eta)},
  \end{split}
\end{equation}%
\endgroup
for the pp part.

%=================================================================%
\section{Algorithm}
\label{sec:algorithm}
%=================================================================%

\begin{figure*}
  \includegraphics[width=0.66\linewidth]{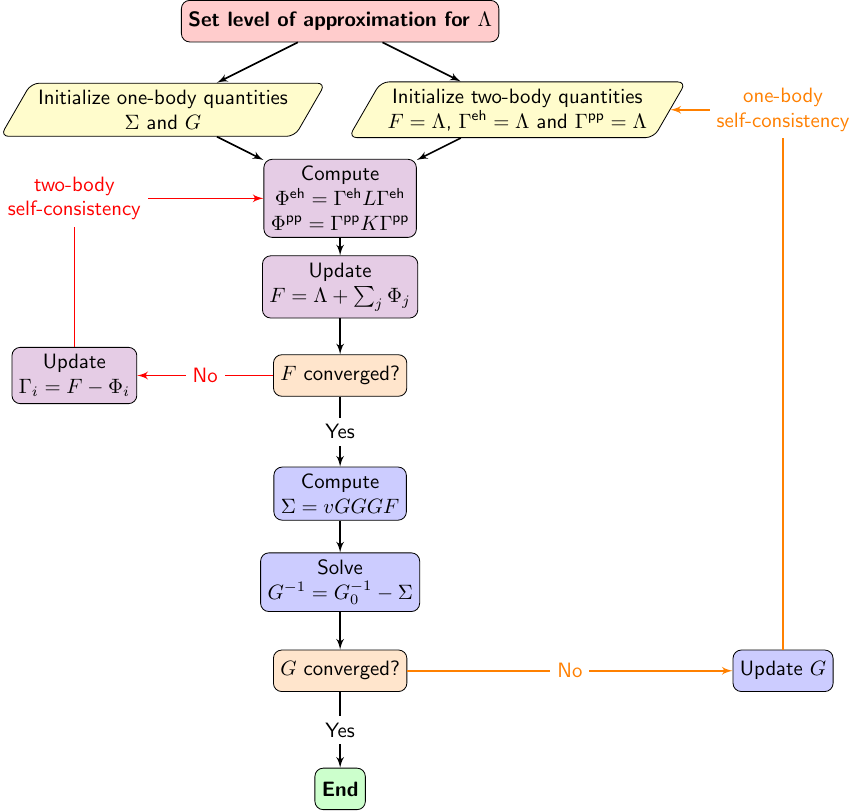}
  \caption{Schematic representation of the self-consistent algorithm used in this work to solve the parquet equations.\label{fig:fig8}}
\end{figure*}

Various equations have been introduced and derived thus far.
While it is clear that they are interdependent, the optimal strategy for achieving self-consistency among them is not obvious.
This section aims to address this question.
Figure~\ref{fig:fig8} presents a schematic pseudo-code to guide the discussion.
As previously mentioned, the first step involves selecting an approximation for the fully irreducible two-body vertex. 
As discussed above, the most common choice is the antisymmetrized bare Coulomb interaction.
Next, all quantities that will be iteratively updated must be initialized.
The reducible vertices are set to zero, and the ``irreducible in one channel only'' vertices are initialized from the fully irreducible vertex.
In this work, the self-energy is always initialized as the HF self-energy, although one could alternatively start from a Kohn-Sham mean-field Hamiltonian, or even from $GW$ quasiparticle energies.

As shown above, the self-energy is expressed only in terms of $\Phi^\eh$ and $\Phi^\pp$.
These reducible vertices are computed using the ``irreducible in one channel only'' vertices, which themselves depend on the $\Phi$'s.
This defines the two-body self-consistent loop depicted in Fig.~\ref{fig:fig8}.
The reducible vertices also depend on the one-body Green's function (through the BSEs).
On the other hand, $G$ is determined using the Dyson equation and depends on the vertex, since the self-energy is calculated from $F$.
This mutual dependence introduces a second level of self-consistency, referred to as one-body self-consistency.

%%%  FIG 8  %%% 

The following list enumerates each step of the self-consistent procedure employed to solve the parquet equations:
\begin{enumerate}
\item Select a starting point for the one-body energies $\epsilon_p$ and electron repulsion integrals $\braket{pq}{rs}$.
\item Initialize $\Phi^\eh = \Phi^\pp = 0$ and $\Gamma^\eh = \Gamma^\pp = \Lambda$.
\item Build the eh and pp effective Hamiltonians using $\epsilon_p$, $\Gamma^\eh$ and $\Gamma^\pp$ [see Eqs.~\eqref{eq:ABeh} and~\eqref{eq:BCDpp}].
\item Diagonalize $\bcal{H}^{\text{eh}}$ and $\bcal{H}^{\text{pp}}$ to obtain the corresponding eigenvalues and eigenvectors [see Eqs.~\eqref{eq:cHeh} and~\eqref{eq:cHpp}].
\item Compute the eh and pp generalized effective integrals by contracting the eigenvectors with $\Gamma^\eh$ and $\Gamma^\pp$ [see Eqs.~\eqref{eq:parquet_eh_screened_integrals} and~\eqref{eq:parquet_pp_screened_integrals}]
\item Evaluate the reducible vertices $\Phi^\eh$ and $\Phi^\pp$ in the static limit using the generalized effective integrals and the two-body eigenvalues [see Eqs.~\eqref{eq:static_eh_reduc_kernel} and~\eqref{eq:static_pp_reduc_kernel}].
\item Update $\Gamma^\eh$ and $\Gamma^\pp$ using the new reducible vertices [see Eqs.~\eqref{eq:update_Gamma_eh} and~\eqref{eq:update_Gamma_pp}].
\item Check for convergence of the reducible vertices: if the change in $\Phi^\eh$ and $\Phi^\pp$ is smaller than the threshold $\tau_\text{2b}$, proceed to the next step.
Otherwise, return to step 3.
\item Build the self-energy from the converged two-body quantities [see Eqs.~\eqref{eq:Sig_2_pq},~\eqref{eq:Sig_eh_pq}, and~\eqref{eq:Sig_pp_pq}] and solve the corresponding quasiparticle equation in the diagonal approximation.
\item Check for convergence of the one-body energies: if the change in $\epsilon_p$ is below the threshold $\tau_\text{1b}$, the procedure is complete.
Otherwise, return to step 1.
\end{enumerate}

In this algorithm, one-body self-consistency is monitored at the level of the quasiparticle energies and the Dyson equation is solved within the diagonal approximation (so only the diagonal elements of the self-energy are required).
This scheme, which we refer to as eigenvalue-only parquet approximation (evPA), is fully analogous to the ev$GW$ partially self-consistent scheme.\cite{Hybertsen_1986,Shishkin_2007a,Blase_2011a,Faber_2011,Marom_2012,Kaplan_2016,Wilhelm_2016,Rangel_2016}
It is common in the $GW$ framework to omit any one-body self-consistency altogether, leading to the well-known one-shot (os) $G_0W_0$ scheme.\cite{Strinati_1980,Hybertsen_1985,Godby_1988,Linden_1988,Northrup_1991,Blase_1994,Rohlfing_1995} 
Its parquet analogue is denoted osPA.
On the other hand, in quasiparticle self-consistent (qs) $GW$, both quasiparticle energies and orbitals are updated self-consistently, while enforcing a static and Hermitian self-energy.\cite{Faleev_2004,vanSchilfgaarde_2006,Kotani_2007,Ke_2011,Gui_2018,Kaplan_2016,Forster_2021,Marie_2023}
We refer to the corresponding parquet analogue as qsPA.
In contrast to osPA and evPA, qsPA requires the full self-energy matrix, and convergence is assessed via the standard DIIS commutator between the effective Fock and density matrices.
For further details on the ev and qs schemes, we refer the reader to Ref.~\onlinecite{Marie_2024a}.

To conclude this section, we would like to comment on the computational cost associated with this algorithm.
If the self-energy is evaluated analytically, the most expensive step in the $G_0W_0$ algorithm is the diagonalization of the eh-RPA problem, which scales as $\order*{K^6}$.
For partially self-consistent schemes, such as ev$GW$ or qs$GW$, the scaling is formally the same but with a larger prefactor due to the multiple iterations.
Similarly, the computational cost of the one-shot pp $T$-matrix algorithm ($G_0T_0^\pp$) is dominated by the pp-RPA diagonalization, which also scales as $\order*{K^6}$.
Of course, the prefactor of the pp-RPA is significantly larger than its eh counterpart as their scaling with respect to the occupied and virtual orbitals are $\order*{(O^2+V^2)^3} = \order*{V^6}$ and $\order*{(2OV)^3} = \order*{O^3V^3}$, respectively.

At each two-body iteration of the parquet formalism, one must diagonalize the eh-BSE and pp-BSE effective Hamiltonians defined in Eqs.~\eqref{eq:cHeh} and~\eqref{eq:cHpp}. 
Each diagonalization entails an overall computational scaling of $\order*{K^6}$.
This results in a larger prefactor than methods based on a single correlation channel, although the overall formal scaling remains unchanged.

On the other hand, the computational cost associated with the construction of the self-energy is larger in the parquet case.
As shown in Eqs.~\eqref{eq:Sig_eh_pq} and~\eqref{eq:Sig_pp_pq}, each element of the eh and pp self-energy contributions can be computed in $\order*{K^5}$ operations.
This results in an overall $\order*{K^6}$ computational cost for the diagonal elements of the self-energy, or $\order*{K^7}$ if all matrix elements are evaluated.
However, by introducing suitable intermediates, this $\order*{K^7}$ cost can be reduced to $\order*{K^6}$, provided one is willing to invest $\order*{K^4}$ in memory storage.
The expression of the self-energy in terms of these intermediates can be found in the \SupInf.
Hence, the formal scaling of the parquet approximation, with the static kernel approximation of Sec.~\ref{sec:frequency_space}, is also $\order*{K^6}$, albeit with a much larger prefactor than for $GW$.
Of course, one should mention that most practical $GW$ implementations scale as $\order*{K^4}$ by leveraging numerical integration of the self-energy and employing techniques such as density fitting.
Adapting these algorithmic strategies to the parquet framework is an important direction for future work.

%=================================================================%
\section{Computational details}
\label{sec:comp_det}
%=================================================================%

The parquet approximation, which was described in detail above, has been implemented in an open-source, in-house program, named \textsc{quack}.\cite{QuAcK}
This implementation was carried out in both spin and spatial orbitals, allowing us to check the correctness of the spin-adaptation performed in the \SupInf.
For osPA and evPA, quasiparticle energies are computed by directly solving the non-linear, frequency-dependent quasiparticle equation, without resorting to linearization.

The convergence of the two-body loop has been accelerated using the DIIS procedure applied to the tensor elements of $\Phi^\eh$ and $\Phi^\pp$.
Additionally, it was found that introducing damping prior to DIIS further enhances convergence.
This observation suggests that the current implementation of DIIS may not be optimal for this set of equations.
For example, although $\Phi^\eh$ and $\Phi^\pp$ are computed and stored in the present implementation, this may not be strictly necessary.
Alternatively, one could extrapolate the effective integrals rather than the reducible vertices.
Further investigation is warranted to determine whether such modifications could improve the robustness and efficiency of the parquet solver.

Our study is limited to closed-shell neutral reference systems, consistently employing a restricted formalism.
Throughout this work, HF orbitals and energies are systematically used as the starting point.
For all calculations in this work, we employ the aug-cc-pVTZ basis set.
The molecular geometries are taken from the \textsc{quest} database.\cite{Loos_2025,Marie_2024}

%=================================================================%
\section{Results}
\label{sec:results}
%=================================================================%

%%%%%%%%%%%%%%%%%%%%%%%%%%%%%%%%%%%%%%%%%%%%%%%%%%
\subsection{The fluctuation-exchange approximation}
\label{sec:FLEX}
%%%%%%%%%%%%%%%%%%%%%%%%%%%%%%%%%%%%%%%%%%%%%%%%%%

\begin{figure}[b!]
  \includegraphics[width=0.9\linewidth]{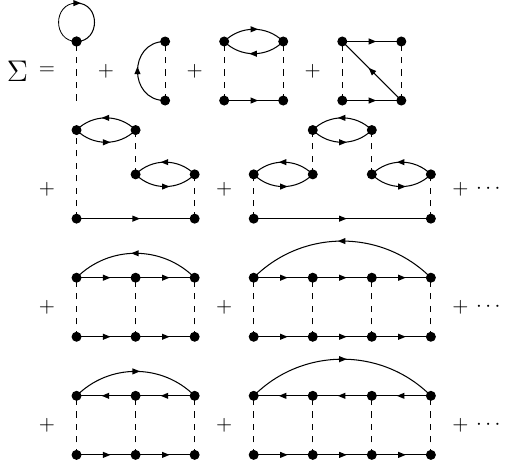}
  \caption{Diagrammatic representation of the FLEX self-energy.\label{fig:fig9}}
\end{figure}

The first step of this study is to investigate the parquet approximation without any self-consistency.
In this case, the self-energy is equal to its second-order part augmented by three resummations of diagrams corresponding to the three two-body correlation channels without any coupling.
This approximation is known as FLEX in the literature and is represented diagrammatically in Fig.~\ref{fig:fig9}.
At the first iteration, the reducible vertices are zero, so the BSE problems reduce to their RPA counterparts.
Consequently, when only one of the three correlation channels is retained, this scheme reduces to the $GW$, the pp $T$-matrix, or the eh $T$-matrix approximation.\cite{Orlando_2023b}
(Strictly speaking, to correctly recover the $GW$ approximation based on the direct eh-RPA problem, one should use only the direct Coulomb term in the fully irreducible vertex $\Lambda$.)
This relation between known self-energy approximations and the first iteration of the parquet approximation has been used as a sanity check for our implementation.

%%%  FIG 9  %%% 
%%% %%% %%% %%%

The performance of the FLEX approximation is evaluated on 8 principal IPs of the 10- and 14-electron series of molecules.
The corresponding results are reported in Table~\ref{tab:tab1}.
Two variants of FLEX have been considered; they are obtained by enforcing, or not, the Tamm-Dancoff approximation (TDA) in the eh and pp eigenvalue problems, \ie, neglecting the coupling blocks in their respective eigenvalue problems.
For both variants, the value of $\eta$ was set to zero.
The $G_0W_0$ and $G_0T_0^\pp$ (without TDA) results are reported for comparison, as well as the full configuration interaction (FCI) reference values extracted from Ref.~\onlinecite{Marie_2024}.

The mean absolute error (MAE) of the FLEX approximations on this small test set is \SI{1.13}{\eV} and \SI{0.66}{\eV} without and with enforcing the TDA, respectively.
These results are definitely worse than those of the single-channel $G_0W_0$ and $G_0T_0^\pp$ approximations.
This error can be attributed to the third class of diagrams that is resummed in the FLEX approximation, namely the eh ladders [see last line of Fig.~\ref{fig:fig9}].
It has been shown in the literature that a self-energy based solely on this class of diagrams performs poorly in molecular systems.\cite{Orlando_2023b}
Note that for the full FLEX scheme, the error associated with the principal IP of \ce{BF} is a large outlier without any regularization and has been removed from the statistics.

%%% TABLE 1 %%%
\begin{table}[h!]
  \caption{Principal IP of 8 small molecules in the aug-cc-pVTZ basis set.
    The FLEX results have been obtained with (FLEX TDA) or without (FLEX RPA) enforcing the TDA.
    The corresponding renormalization factors (or spectral weights) are reported in parentheses.\label{tab:tab1}}
  \begin{ruledtabular}
    \begin{tabular}{lccccc}
      Molecule & \mcc{FCI} & \mcc{FLEX RPA} & \mcc{FLEX TDA} & \mcc{$G_0W_0$} & \mcc{$G_0T_0^\pp$} \\
      \hline
      \ce{Ne}  & 21.46 & 20.04(0.83) & 20.41(0.86) & 21.43 & 21.08 \\
      \ce{HF}  & 16.15 & 14.40(0.75) & 14.85(0.79) & 16.24 & 15.72 \\
      \ce{H2O} & 12.68 & 11.25(0.70) & 11.54(0.76) & 12.88 & 12.36 \\
      \ce{NH3} & 10.90 & 10.23(0.69) & 10.22(0.75) & 11.20 & 10.72 \\
      \ce{CH4} & 14.38 & 14.90(0.79) & 14.37(0.83) & 14.75 & 14.28 \\
      \ce{BF}  & 11.15 &             & 11.62(0.73) & 11.33 & 10.96 \\
      \ce{CO}  & 13.93 & 15.24(0.58) & 14.02(0.73) & 14.78 & 14.32 \\
      \ce{N2}  & 15.49 & 14.71(0.64) & 14.93(0.72) & 16.35 & 15.72 \\
      \hline
      MSE      &       &    -0.60    &    -0.52    &  0.35 & -0.12 \\
      MAE      &       &   \m1.13    &   \m0.66    &  0.36 &  0.28 \\
    \end{tabular}
  \end{ruledtabular}
\end{table}
%%% %%% %%% %%% 

%%%%%%%%%%%%%%%%%%%%%%%%%%%%%%%%%%%%%%%%%%%%%%%%%%
\subsection{Two-body self-consistency}
\label{sec:2b_self_consistency}
%%%%%%%%%%%%%%%%%%%%%%%%%%%%%%%%%%%%%%%%%%%%%%%%%%

The previous results have highlighted the need to properly couple the two-body correlation channels if they are considered simultaneously.
This section will go beyond the FLEX approximation by investigating the coupling of channels induced by the two-body self-consistency.
Note that the one-body self-consistency is not considered at this stage.
The two-body self-consistency is performed until the absolute change in $\Phi^\eh$ and $\Phi^\pp$ is smaller than $\tau_\text{2b} = 10^{-4}$. 
Then, the corresponding self-energy and one-body energies are obtained (under the diagonal approximation and with $\eta=0$), and the algorithm is stopped after this single one-body iteration.
This scheme was denoted as osPA in Sec.~\ref{sec:algorithm}.

The comparison of osPA with FLEX allows us to gauge the impact of the two-body self-consistency.
Unfortunately, in practice, the implementation described in Sec.~\ref{sec:algorithm} is hindered by two challenges.
First, the self-consistent update of the reducible vertices was found to diverge in most cases.
This can be prevented through regularization, \ie, by using a finite value of $\eta$ to compute $\Phi^\eh$ and $\Phi^\pp$ [see Eqs.~\eqref{eq:static_eh_reduc_kernel} and~\eqref{eq:static_pp_reduc_kernel}].
In this work, the $\ii\eta$ regularizer has been replaced by a more effective energy-dependent regularizer\cite{Evangelista_2014b,Monino_2022,Marie_2023}
\begin{subequations}
  \begin{align}
    \begin{split}
      \ii (\Phi^{\eh})_{pqrs} = &- \sum_{n} \frac{M^{\eh}_{pr,n} M^{\eh,*}_{sq,n}}{\Omega_n^\eh} (1 - e^{-2s_{\text{2b}}(\Omega_n^\eh)^2}) \\
                                &- \sum_{n} \frac{M^{\eh,*}_{rp,n} M^{\eh,\text{d}}_{qs,n}}{\Omega_n^\eh} (1 - e^{-2s_{\text{2b}}(\Omega_n^\eh)^2}), 
    \end{split} \\
    \begin{split}
      \ii (\Phi^{\pp}_{\text{P}})_{pqrs} = &- \sum_{m} \frac{M^{\ee}_{pq,m} M^{\ee,*}_{rs,m}}{\Omega_m^{\ee}} (1 - e^{-2s_{\text{2b}}(\Omega_m^\ee)^2}) \\
                                           &+ \sum_{m} \frac{M^{\hh,*}_{pq,m} M^{\hh}_{rs,m}}{\Omega_m^{\hh}} (1 - e^{-2s_{\text{2b}}(\Omega_m^\hh)^2}).
    \end{split}
  \end{align}
\end{subequations}
Indeed, it was shown in Ref.~\onlinecite{Marie_2023}, for example, that such an energy-dependent regularization greatly improves the convergence properties of the qs$GW$ scheme with respect to the energy-independent regularizer $\ii\eta$.

Similarly, in the context of the parquet approximation, regularization was found to be essential for achieving convergence. 
However, it does not resolve all issues.
In particular, we have observed that triplet instabilities\red{, manifesting as complex eigenvalues,} can arise in the eh-BSE eigenvalue problem during the self-consistent cycle.
This problem further plagues the convergence but can be alleviated, or at least avoided, by enforcing the TDA.

\begin{figure*}[t!]
  \includegraphics[width=0.4\linewidth]{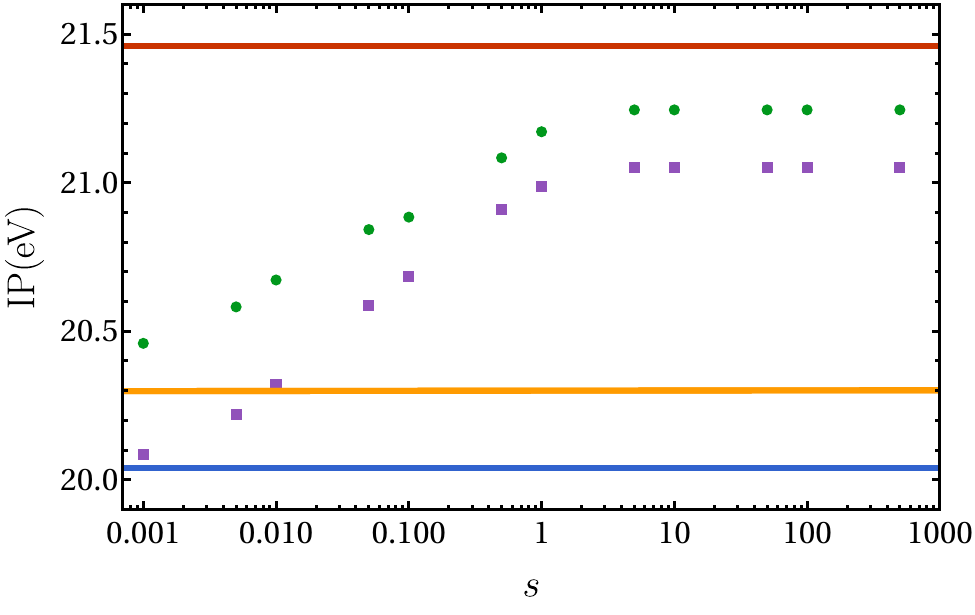}
  \includegraphics[width=0.4\linewidth]{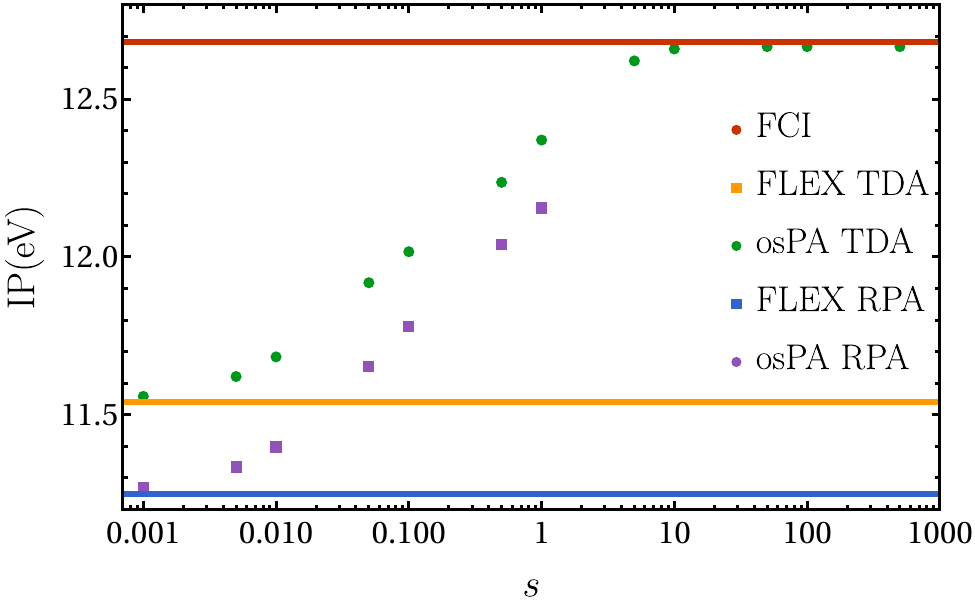}
  \caption{Principal IP of \ce{Ne} and \ce{H2O} in the aug-cc-pVTZ basis computed with osPA, with or without TDA, as a function of the regularization parameter $s_{\text{2b}}$.
    The FCI reference value and the FLEX values are reported for comparison.\label{fig:fig10}}
\end{figure*}

Figure~\ref{fig:fig10} displays the principal IP for \ce{Ne} and \ce{H2O} in the aug-cc-pVTZ basis set obtained after this first one-body iteration as a function of the regularization parameter $s_{\text{2b}}$.
For $s_{\text{2b}}=0$, the reducible vertices vanish due to complete regularization, and the parquet approximation reduces to the FLEX approximation.
The left panel shows that in both cases, \ie, with or without TDA, the results are improving when the flow parameter is increased until a plateau is reached around $s_{\text{2b}}=10$.
This evidences the fact that the two-body correlation channels are getting renormalized through their mutual coupling.
For the case of \ce{H2O}, a similar behavior is observed for the TDA results.
However, we could not achieve convergence for $s_{\text{2b}} > 1$ without TDA, due to triplet instabilities in the eh-BSE eigenvalue problem that arise after only a few iterations.

%%%  FIG 10  %%% 
%%% %%%  %%% %%%

Table~\ref{tab:tab2} reports the MSE and MAE for the principal IP of 8 small molecules in the aug-cc-pVTZ basis set as a function of the two-body regularization parameter $s_{\text{2b}}$.
This demonstrates the beneficial impact of two-body self-consistency as the MAE is decreasing when $s$ is increasing.
At $s_{\text{2b}}=100$, the largest value for which every IPs could be converged while enforcing the TDA, the MAE is \SI{0.23}{\eV}.
This MAE is slightly better than the $G_0W_0$ and $G_0T_0^\pp$ ones (see Table~\ref{tab:tab1}).
However, one should keep in mind that this test set is really small and only proper statistics on a large test set will assess the performance of osPA with respect to $G_0W_0$ and $G_0T_0^\pp$.
Interestingly, the MSE of the multi-channel osPA (\SI{-0.04}{\eV}) is in between the MSEs of the single-channel $G_0W_0$ and $G_0T_0^\pp$ methods (\SI{0.35}{\eV} and \SI{-0.12}{\eV}, respectively).
The two-body self-consistency also has an impact on the renormalization factors (or spectral weights).
For example, at $s_{\text{2b}}=100$, every IP considered here has a well-defined quasiparticle character with $Z>0.9$.
This is not the case at the FLEX level, where most renormalization factors are in between $0.6$ and $0.8$.

%%% TABLE 2 %%%
\begin{table}[h!]
  \caption{Principal IP of 8 small molecules in the aug-cc-pVTZ basis computed with osPA, enforcing the TDA, as a function of the regularization parameter $s_{\text{2b}}$.
    The corresponding renormalization factors (or spectral weights) are reported in parentheses.\label{tab:tab2}}
  \begin{ruledtabular}
    \begin{tabular}{lcccccc}
    & \mc{6}{c}{$s_{\text{2b}}$} \\
      \cline{2-7} 
      Molecule & \mcc{$0.001$} & \mcc{$0.01$} & \mcc{$0.1$} & \mcc{$1$} & \mcc{$10$} & \mcc{$100$} \\
      \hline
      \ce{Ne}  & 20.46 & 20.67 & 20.88 & 21.17 & 21.24 & 21.24(0.94) \\
      \ce{HF}  & 14.88 & 15.09 & 15.45 & 15.83 & 16.05 & 16.05(0.92) \\
      \ce{H2O} & 11.56 & 11.68 & 12.02 & 12.37 & 12.66 & 12.67(0.92) \\
      \ce{NH3} & 10.22 & 10.26 & 10.39 & 10.60 & 10.84 & 10.85(0.93) \\
      \ce{CH4} & 14.36 & 14.29 & 14.09 & 14.03 & 14.11 & 14.11(0.93) \\
      \ce{BF}  & 11.61 & 11.53 & 11.17 & 10.62 & 10.68 & 10.70(0.95) \\
      \ce{CO}  & 14.02 & 14.00 & 13.95 & 14.02 & 14.21 & 14.22(0.93) \\
      \ce{N2}  & 14.94 & 15.00 & 15.28 & 15.72 & 15.94 & 15.94(0.94) \\
      \hline                    
      MSE      & -0.51 & -0.45 & -0.36 & -0.22 & -0.05 & -0.04 \\
      MAE      &\m0.65 &\m0.56 &\m0.37 &\m0.30 &\m0.23 &\m0.23 \\
    \end{tabular}
  \end{ruledtabular}
\end{table}
%%% %%% %%% %%% 

%%%%%%%%%%%%%%%%%%%%%%%%%%%%%%%%%%%%%%%%%%%%%%%%%%
\subsection{Full self-consistency}
\label{sec:}
%%%%%%%%%%%%%%%%%%%%%%%%%%%%%%%%%%%%%%%%%%%%%%%%%%

Finally, the full self-consistency, \ie, at both the one- and two-body levels, is investigated.
The one-body self-consistency has been implemented in two different ways.
First, it is performed in the diagonal approximation and it is thus analogous to the ev$GW$ scheme, as discussed in Sec.~\ref{sec:algorithm}.
It is referred to as evPA.
Once again, an energy-dependent regularization has been employed to facilitate convergence.
For example, the third term in Eq.~\eqref{eq:Sig_eh_pq}, for a diagonal matrix element evaluated at $\omega = \epsilon_p$, has been transformed into
\begin{equation}
  \label{eq:srg_self_energy}
  \sum_{ijan} \frac{\mel{pa}{}{ij}}{\Delta_{paij}} (1 - e^{-2s_{\text{1b}}\Delta_{paij}^2})  \frac{M^{\eh,*}_{ai,n} M^{\eh}_{jq,n}}{\Delta_{ain}} (1 - e^{-2s_{\text{1b}}\Delta_{ain}^2}),
\end{equation}
where $\Delta_{paij} = \epsilon_p +  \epsilon_a - \epsilon_i - \epsilon_j$ and $\Delta_{ain} = \epsilon_a - \epsilon_i + \Omega_{n}$.

The equivalent of qs$GW$, denoted as qsPA, has also been implemented.
In this scheme, the frequency-dependent self-energy is approximated by a static Hermitian self-energy and is diagonalized at each one-body iteration to obtain the quasiparticle energies.
The static parquet self-energy that has been used in this study is inspired by the qs$GW$ self-energy of Ref.~\onlinecite{Marie_2023} and is reported in the \SupInf.
It also relies on an energy-dependent regularization as in Eq.~\eqref{eq:srg_self_energy}.

The 8 principal IPs of the small set considered so far have been obtained with, and without, one-body self-consistency.
The corresponding results are given in Table~\ref{tab:tab3}.
The one-body convergence threshold $\tau_\text{1b}$ was set to $10^{-4}$, while the two-body convergence threshold was still set to $\tau_\text{2b} = 10^{-4}$.
The two-body regularization parameter has been set to $s_\text{2b} = 50$ based on the results of the previous section.
The impact of the one-body regularization parameter on evPA and qsPA has also been investigated, and it was found that $s_\text{1b} = 50$ is the largest value ensuring convergence.
Note that, for \ce{CO} and \ce{N2}, $s_\text{1b}$ was decreased to 25 and 10, respectively, to reach convergence at the evPA level.

Table~\ref{tab:tab3} shows that the impact of one-body self-consistency at the qsPA level is to decrease the IPs by \SIrange{0.1}{0.3}{\eV} approximately.
Unfortunately, the MSE without one-body self-consistency was already negative, so this leads to a slightly worse MAE, with respect to osPA, of \SI{0.26}{\eV}.
The effect of the evPA one-body self-consistency is also to decrease IPs on average, yielding a MAE of \SI{0.35}{\eV}.
This set of results suggests that the one-body self-consistency has only a minor impact on the principal IPs compared to the two-body self-consistency.
However, some early tests reveal that this has a much larger influence on the two-body excitations provided by the eh- and pp-BSE.

%%% TABLE 3 %%%
\begin{table}[h!]
  \caption{Principal IP of 8 small molecules in the aug-cc-pVTZ basis computed at various self-consistent schemes within the parquet and $GW$ approximations. 
  In the parquet approximation, the TDA is enforced and the regularization factors are $s_\text{1b} = 50$ and $s_\text{2b} = 50$.\label{tab:tab3}}
  \begin{ruledtabular}
  \begin{tabular}{lccccccc}
    Molecule &  osPA &  evPA &  qsPA & $G_0W_0$ & ev$GW$ & qs$GW$ \\
    \hline
    \ce{Ne}  & 21.28 & 21.01 & 21.28 & 21.43 & 21.25 & 21.59 \\
    \ce{HF}  & 16.06 & 15.79 & 15.98 & 16.24 & 16.07 & 16.35 \\
    \ce{H2O} & 12.68 & 12.51 & 12.53 & 12.88 & 12.78 & 12.88 \\
    \ce{NH3} & 10.89 & 10.77 & 10.65 & 11.20 & 11.16 & 11.09 \\
    \ce{CH4} & 14.14 & 14.01 & 13.83 & 14.75 & 14.74 & 14.62 \\
    \ce{BF}  & 10.71 & 10.66 & 10.56 & 11.33 & 11.35 & 11.17 \\
    \ce{CO}  & 14.23 & 14.14 & 13.76 & 14.78 & 14.74 & 14.32 \\
    \ce{N2}  & 15.96 & 16.11 & 15.44 & 16.35 & 16.27 & 15.91 \\
    \hline                    
    MSE      & -0.02 & -0.14 & -0.26 &  0.35 & 0.28 & 0.23 \\
    MAE      &\m0.22 &\m0.35 &\m0.26 &  0.36 & 0.35 & 0.23 \\
  \end{tabular}
\end{ruledtabular}
\end{table}
%%% %%% %%% %%%

%%%  FIG 11  %%% 
\begin{figure*}%[t!]
  \includegraphics[width=\linewidth]{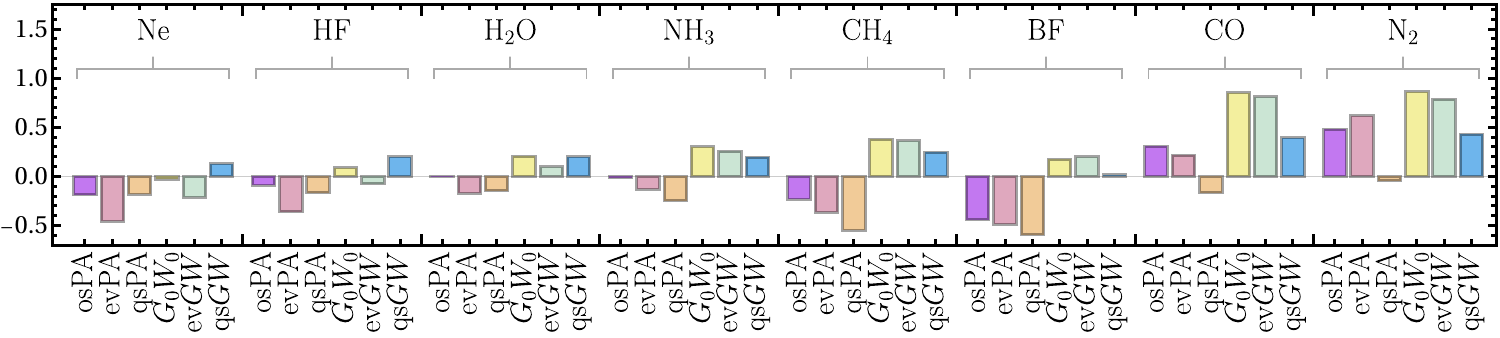}
  \caption{Histogram of errors, with respect to FCI, of the principal IP of 8 small molecules in the aug-cc-pVTZ basis computed at various self-consistent levels within the parquet and $GW$ approximations. 
  In the parquet approximation, the TDA is enforced and the regularization factors are $s_\text{1b} = 50$ and $s_\text{2b} = 50$.\label{fig:fig11}}
\end{figure*}
%%% %%% %%% %%%

%=================================================================%
\section{Conclusion}
\label{sec:conclusion}
%=================================================================%

In this work, we have presented the first implementation and assessment of the parquet formalism for molecular systems, focusing on principal IPs in 10- and 14-electron molecules.
Unlike single-channel approaches such as $GW$ or the pp $T$-matrix, the parquet framework treats all two-body correlation channels on equal footing, preserving crossing symmetry and Pauli exclusion principle.
This implementation relies on the common parquet approximation for the fully irreducible two-body vertex as well as an additional static kernel approximation analogous to the usual eh-BSE formalism.

Our results highlight both the opportunities and challenges of this approach. 
At the lowest level of self-consistency, the FLEX approximation performs worse than single-channel methods due to the poor behavior of the eh-ladder diagrams. 
However, introducing two-body self-consistency significantly improves the results: IPs are brought to within \SI{0.3}{\eV} of FCI reference values, with quasiparticle weights close to unity. 
This demonstrates that a proper coupling of correlation channels is essential and that their mutual renormalization plays a central role in accuracy.

We also explored one-body self-consistency via two schemes. 
In the evPA approach, the diagonal approximation of the self-energy is enforced, and only quasiparticle energies are updated self-consistently. 
In the qsPA scheme, the full self-energy matrix is computed, and both quasiparticle energies and Dyson orbitals are determined self-consistently using a static and Hermitian version of the self-energy. 
Unlike two-body self-consistency, one-body self-consistency has only a modest effect, which does not seem to improve IPs.
However, preliminary evidence suggests it may be more consequential for two-body excitations, a direction we leave for future work.

From a computational perspective, the parquet approximation scales formally as $\order{K^6}$, with a larger prefactor than $GW$ due to the simultaneous treatment of multiple correlation channels. 
Unlike $GW$, however, it delivers a broader range of observables, providing access not only to IPs and EAs, but also to neutral (optical) excitations as well as DIPs and DEAs.
Convergence issues necessitated the use of energy-dependent regularizer instead of the usual imaginary shift, which proved crucial for stabilizing the one- and two-body self-consistency.

Overall, this study shows that parquet theory offers a promising route to go beyond $GW$ in molecular systems. 
Its balanced treatment of correlation channels makes it particularly appealing for situations where $GW$ is known to struggle, such as strongly correlated regimes or cases where multiple scattering mechanisms compete. 
At the same time, our findings underline the importance of algorithmic advances to reduce computational cost and improve convergence. 
Future work will focus on extending the present implementation beyond the static kernel approximation, exploring the impact of parquet self-consistency on excitation spectra and total energies, investigating the properties (such as positive semi-definiteness) of the parquet self-energy, \cite{Bruneval_2025} and integrating modern low-scaling strategies to broaden applicability.

%=================================================================%
\section*{Supplementary Material}
\label{sec:supmat}
%=================================================================%

See the \SupInf for detailed derivations of all equations, their projection into the spin-orbital basis, and their spin-adaptation, complementing the results presented in the main manuscript.

%=================================================================%
\acknowledgments{This project has received funding from the European Research Council (ERC) under the European Union's Horizon 2020 research and innovation programme (Grant agreement No.~863481).}
%=================================================================%

%=================================================================%
\section*{Data availability statement}
%=================================================================%

The data that support the findings of this study are available within the article and its supplementary material.

%=================================================================%
\section*{References}
%=================================================================%

%=================================================================%
\bibliography{parquet.bib}
%=================================================================%

\end{document}